\RequirePackage{fix-cm}
\documentclass[smallextended]{svjour3}       
\smartqed  
\usepackage{graphicx}
\usepackage{color}
\usepackage{ulem}
\begin{document}

\title{Are rare rogue fluctuations generic to strongly nonlinear and non-integrable systems?
}

\titlerunning{Rogue fluctuations}        

\author{Rahul Kashyap         \and
        Surajit Sen 
}


\institute{Rahul Kashyap  \at
              Physics Department, State University of New York at Buffalo, Buffalo, New York 14260-1500, USA \\
              Tel.: +1 716 645 2017\\
              Fax: +1  716 645 2507\\
             \email{rahulkashyap7557@gmail.com}           
           \and
           Surajit Sen \at
            Physics Department, State University of New York at Buffalo, Buffalo, New York 14260-1500, USA \\ 
            Tel: +1 716 907 4961\\
            Fax: +1 716 645 2507\\
            \email{sen@buffalo.edu}\\  
}

\date{Received: date / Accepted: date}

\maketitle

\begin{abstract}
Extensive dynamical simulations are used to explore the possible existence of sudden sufficiently large energy or rogue fluctuations (RF) at late times and across short time windows in the {\it strongly nonlinear regime} of the $\beta$-Fermi-Pasta-Ulam-Tsingou (FPUT) type systems. Our studies build on a study of RF in the non-dissipative granular chain system and suggest that {\it rare RF may be generic to non-integrable, strongly nonlinear systems at late enough times}. We comment on the role of initial conditions and the intriguing influence of harmonic forces on these strongly nonlinear systems. The RF under focus here are distinct from the well known Peregrine solitons used to describe rogue waves via the weakly nonlinear Schr\"odinger equation.

\keywords{Fermi-Pasta-Ulam-Tsingou system \and Quasi-equilibrium \and Nonlinear Systems}
\end{abstract}

\section{Introduction}
\label{intro}
When exploring many particle systems with nonlinear interactions, we often learn about continuum nonlinear equations of motion that are integrable \cite{Re03,Ja91,Wh74} and yield solutions in the form of solitons or localized excitations such as breathers and/or bound solitons (see for example in \cite{Ch75}). Among these we have the Korteweg deVries equation, the nonlinear Schr\"odinger equation and others.  Then there is an exactly solvable discrete system, the  monodispersed Toda lattice \cite{To70,To81,Fe82}. A feature of the Toda system is that the solitary waves (SWs) typically suffer a phase shift when they interact with one another. Other than the phase shift they remain unscathed \cite{To81,Re03,Fo92,Sh14}. However, the vast majority of equations of motion for many body systems with nonlinear forces are not necessarily accurately represented by integrable equations of motion.  Hence there is a knowledge gap that needs to be addressed by examining the dynamics of nonintegrable, strongly nonlinear systems at both early and late times.

It turns out that the study of SW interactions with each other and in systems with boundaries offer insights into the consequences of non-integrability. Further, this is a mature subject that is well over 40 years old \cite{Ab76,Ei77,Bo80,Na81,Ko87,Se01,De20}. For these non-integrable systems one often finds that SWs of finite spatial expanse, localized excitations and acoustic-like oscillations  are the energy carriers \cite{Bi78,Ka17}. Further, one may encounter unknown (neither SWs nor localized excitations but some combination of both) and unstable nonlinear objects that are neither of the three just noted \cite{Ka17,Fl89,Se03,KM09,Av11}. These objects interact in nontrivial ways \cite{Ma00,Ma02,Jo05,Sa11,Ka19}. 

The SW-SW and SW-wall interactions may {\it{eventually}} drive the system to an equilibrium state. The equilibrium state is characterized by Maxwell's Gaussian distribution of velocities and the system's energy is approximately equally distributed among the available degrees of freedom thereby leading to equipartitioning of energy and the natural introduction of an equilibrium temperature of the system \cite{Pa73,Li00}. Additionally, in equilibrium there is no memory of the initial perturbation that was used to remove the system from the original equilibrium state and the system is presumed to be ergodic \cite{Lo86,Le82,Le83,Se91,Se06,Le07}. The approach to some form of equilibrium-like state of an interacting many body system is a subject of considerable fundamental interest in the context of the work presented below \cite {Ki46,Zw65,Be66,Be71,Ku2}.


In a recent study \cite{Ha14} we pushed the idea of strongly perturbing a system and considered an intrinsically nonlinear many particle system represented by an alignment of elastic beads in gentle contact interacting via the one-sided Hertz potential (i.e., no interaction upon contact breaking) and held between two fixed end walls \cite{He81}. We ignored the role of dissipative losses in our studies and considered a conservative system. We gave each bead a random velocity at initiation. These {\it strong} perturbations led to the development of an early phase of the system that may be viewed as one with many interacting SWs. Such a problem would be hard to approach analytically. The equations of motion were hence carefully and numerically integrated forward in time to probe the time evolution of the system. The results of the high accuracy simulations were as follows.

It was found that due to the persistent interactions between the SWs, the system ended up for extended times in a phase that was characterized by large kinetic energy fluctuations. These large kinetic energy fluctuations manifested themselves as what we called hotspots (HS) and rogue fluctuations (RF) \cite{Ka17,Ha14}. The state of the system with large kinetic energy fluctuations has been earlier referred to as the quasi-equilibrium (QEQ) state \cite{Se04,KM05,Se05,Av09}. Upon continued simulation one would expect, based on our earlier work \cite{Ka19,Fu20,Pr17b,Pr17a}, that the system would eventually reach a state with energy equipartitioning and hence an equilibrium state.

It is worth noting that ideas akin to that of the QEQ state appear to have been independently developed in studies of small quantum systems and are often referred to as the prethermalization phase \cite{Be04,Gr12}. In this context, we ask {\it if RF are generic to nonlinear many-body systems and whether they play a role in the dynamics of the system in QEQ}. In seeking to answer these questions we examine the nature of HS and RF in the $\beta$-Fermi-Pasta-Ulam-Tsingou ($\beta$-FPUT) system and this work is described below \cite{Fe55}. 

In closing here, we should mention that formalisms such as Kubo's linear response theory \cite{Ku86,Se06} have been very successful in describing relaxation processes in a great many systems to the equipartitioned state. However, it is our understanding that applying such an approach to study relaxation in these strongly perturbed systems in an analytic manner is still not practical and this is why the current work is based purely on dynamical simulations based on the well tested and openly accessible PULSEDYN \cite{Ka19} code.

The paper is organized as follows: the model and the simulational details are addressed in Sec. 2, the results are discussed in Sec. 3 and the conclusion and discussions are presented in Sec. 4. Sec 3 is split into 4 subsections: 3.1 and 3.2 which discuss the HSs and the more nuanced behavior of the RF in the $\beta$-FPUT system, 3.3 addresses the studies when different initial conditions are used and 3.4, where we contend that the RF enter late in the QEQ phase.

\section{Model System and Simulational Details}
\label{sec:2}

We consider systems with the $\beta$-FPUT {\it like} Hamiltonian below,
\begin{eqnarray}
H & = & \sum_{i=1}^{N}\frac{p_i^2}{2m}+\sum_{i=1}^{N-1} V(x_i-x_{i+1}),
\end{eqnarray}
where the potential $V(x_i -x_{i + 1})$ is given by \cite{Fe55}
\begin{eqnarray}
V(x_i - x_{i + 1}) & = & \alpha (x_i - x_{i + 1})^2 + \beta (x_i - x_{i + 1})^{2n}.\\ \nonumber
\end{eqnarray} 
Here, $p_i$ and $m_i$ are the momentum and mass of the $i$th particle, respectively, $x_i$ is the displacement of the $i$th particle, $\alpha$ controls the strength of the harmonic term and $\beta$ controls the strength of the nonlinear term in the potential. We control the exponent of the nonlinear potential term by varying $n$. We use $n = 2, 3, 4, 5, 6$ and $7$. Increasing $n$ beyond 7 proves to be too expensive computationally. $N$ is the system size. For the studies reported here we set $N = 100$, which is sufficiently large to observe RF without making the simulations too expensive. 

It is important to note that with the current model we are unable to explore potentials with nonlinear power less than 4 and hence the results shown here cannot be easily connected to those seen in the granular chain system where the nonlinearity in the potential is $5/2$ \cite{Ha14}. We also do not address the consequences of asymmetry of the potential in the realization of RF here, which could have interesting consequences, and would be addressed in future work.

To integrate the force equations, we use the velocity Verlet algorithm \cite{Ka19,Ve67,Al87,Ha92}. The issue of error accumulation over long simulation times sets limits on the time step of integration and the extent of nonlinearity we can consider. We observe that increasing $n$ increases the computational expense and the error associated with the dynamical simulations. For this reason, we use $n = 2, 3$ and 4 for most of our simulations, while $n = 5, 6$ and $7$ are used less often. We have used a time step $\delta t = 10^{-5}$. With this time step, we achieve energy conservation of up to 1 part in $10^9$ on average per time step across extended times. 

\begin{figure*}[tbp]
\centering
\includegraphics[scale=0.45]{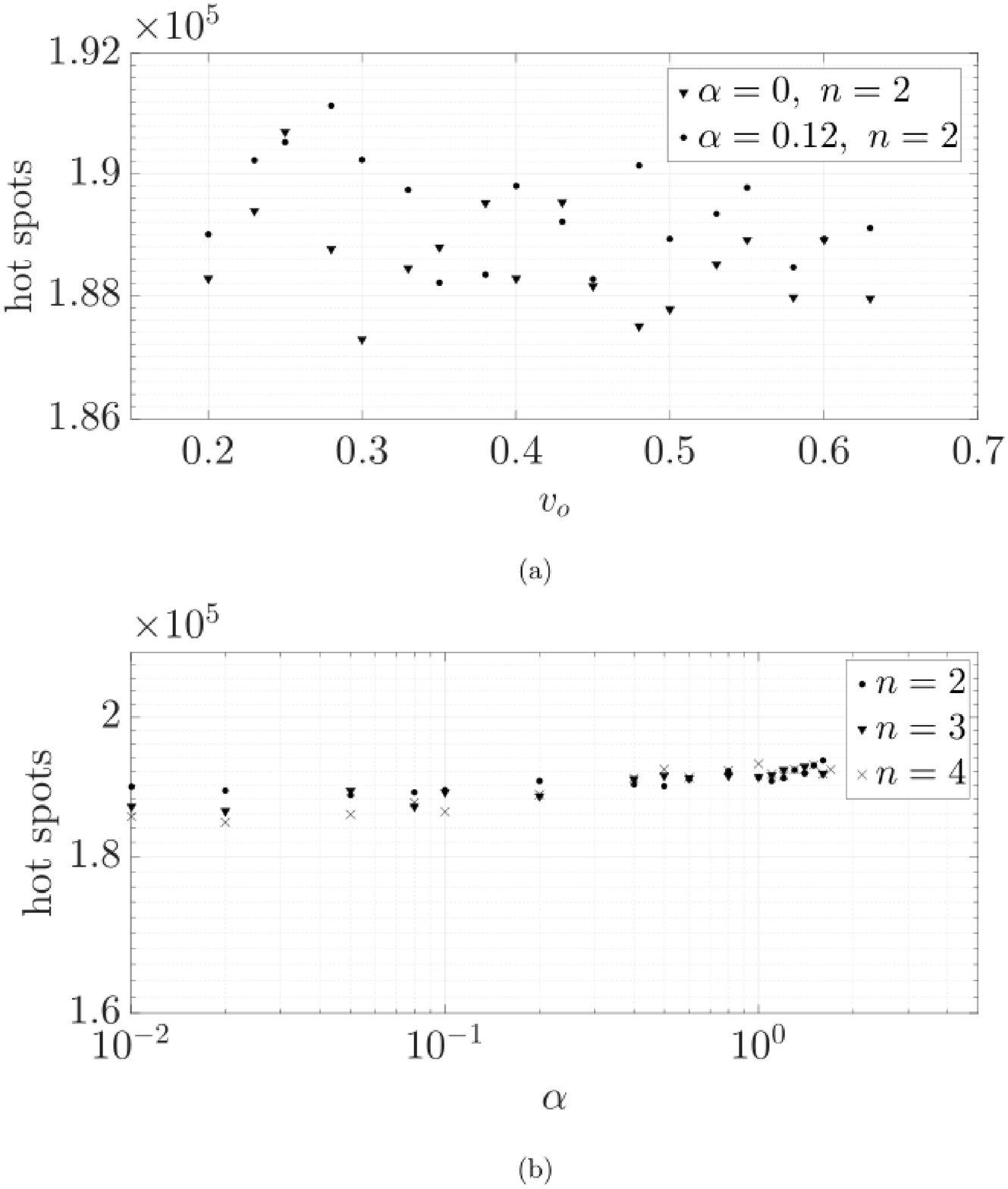}
\caption{Subfigure (a) shows the of number of HS as a function of $v_o$ for two different values of $\alpha$. Subfigure (b) shows the plot of HS as a function of $\alpha$ for $n = 2, 3$ and $4$. }
\label{hotspotsVsVoAlpha}
\end{figure*}

We record data every $1/\delta t$ time steps and the simulations are run to $10^{11}$ time steps, i.e. $N_t = 10^6$, where $N_t$ is the number of recorded snapshots of the system in time. From our simulations, we find that $N_t = 10^6$ is a long enough simulation time such that we recover results for RF that are not influenced by the run times. This length of run time is also computationally feasible. 

While it may be possible to observe RF from any velocity perturbation as an initial condition, simpler the condition or weaker the magnitude of the perturbation, longer it would likely take the RF to form and weaker they are likely to be. Hence, from an intuitive standpoint it makes sense to start from as disordered a state as possible to observe a large number of RF within reasonable times after initiating the system. With this in mind we assign {\it uniformly} distributed random velocity perturbations to each particle within the bounds $v_o$ and $-v_o$. All the results shown in this paper have been obtained with $v_o = 0.6$, though studies with a range of values of $v_o$ have been done. Values much larger than 0.6 are not recommended as they can incur unnecessary calculational errors. While we have also used the {\it Gaussian} and beta distributions to explore the role of initial conditions, all results shown are obtained using the uniform distribution unless stated otherwise. As we shall see in Sec. 3.3, the details of the distributions do not influence the statistics of RF we observe in our studies. With the chosen $v_o$ and $N$ values, the system has a total energy in our dimensionless units of $\sim 10^{-2}$ and relaxes to the early stage QEQ phase within the first few hundred recorded time steps. The relaxation process of the system to QEQ and in QEQ itself is discussed in more detail in Section 3.4.

\section{Results}
\label{results}

We discuss the dynamical simulation based observation of HS and RF for our system below. For our studies, the kinetic energy fluctuations in the system turn out to be important to understand. We will define the kinetic energy per particle as,
\begin{eqnarray}
\delta_K & = & \sqrt{\frac{1}{N_t(N-1)}\sum_{i=1} ^{N}\sum_{j=1} ^{N_t}[E_{K_i}(j) - \frac{\langle E_K\rangle}{N}]^2}.
\end{eqnarray}
Here, $\langle E_K \rangle$ and $ \langle E_K \rangle/N$ denote the average kinetic energy of the system and the average kinetic energy per particle in the QEQ phase, respectively. By the virial theorem, for $\alpha = 0$ in Eq. (2), $\langle E_K \rangle = \frac{2nE}{2n+2}$, where $E$ is the total system energy. 


To identify high energy regions in the chain, we search for sites in the space-time lattice with kinetic energy greater than $\frac{\langle{E_K}\rangle}{N} + 6\delta_K$. We call these sites HS. However, since HS are typically fleeting, we associate RF with fluctuations that last across a small window of time. Therefore, we define a RF as a set of \textit{contiguous} HS on the space time lattice. Here we have set the minimum number of contiguous spots to be 6. The criteria for choosing 6 as the threshold is arbitrary and is discussed below. A larger number than 6 would reduce the number of RF seen while a smaller number would increase the same without significantly affecting the findings reported here. In our experiments, varying this threshold does not affect the trends in any significant manner. We report the results of our calculations in Sec. \ref{results} below.

\subsection{Hotspots}
In all of our simulations, we find that HS are abundant at sufficiently late times. This observation is fully consistent with the findings on the Hertz system reported earlier \cite{Ha14}. To the extent we can accurately study these systems up to late times, we note that the number of HS do not depend in any significant manner on the total energy of the system (set by $v_o$), on the nonlinearity of the system which is controlled by $n$, and the strength of the harmonic forces which is set by $\alpha$ in Eq. (1) (see Fig. \ref{hotspotsVsVoAlpha}). Typical variations in the number of HS seen is $\sim 5-10$\%. The calculations reported here have been run significantly deeper into QEQ and for a system which is 1/5$^{th}$ of the system size compared to that reported in Ref. \cite{Ha14}. This is why the typical number of HS is larger by a factor of 10 in the studies reported here compared to that reported in Ref. \cite{Ha14}.

\begin{figure}[tbp]
\centering
\includegraphics[scale=0.275]{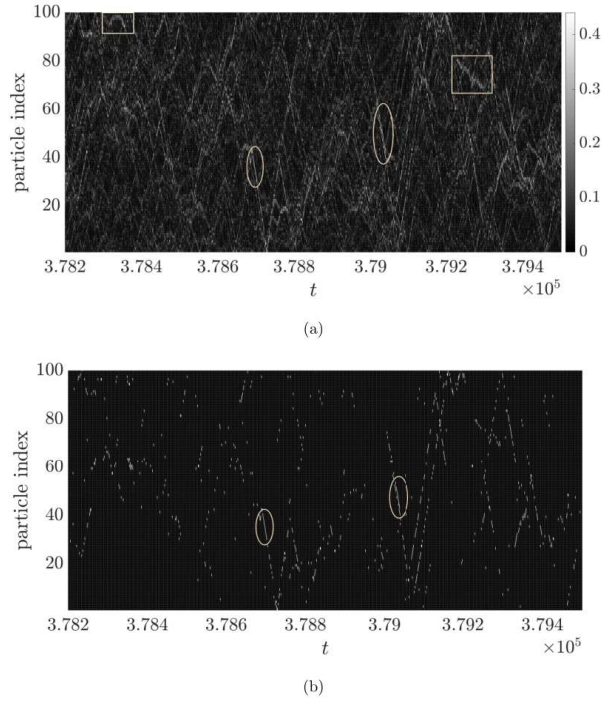}
\caption{Subfigure (a) shows the kinetic energy contour plot on the space-time lattice. The RF are circled in ellipses and the excitations which tend to show localization are shown in the rectangular boxes. Subfigure (b) shows the contour plot of HS filtered from the kinetic energy data. The RF are circled in ellipses. Here, $\alpha = 0$ in Eq. (2) and $n = 2$. Note that not all the RF have been identified on the plots.}
\label{contourMaps}
\end{figure}

\subsection{Rogue Fluctuations}\label{rfResults}

We now turn our attention to the nature of RF in the $\beta$-FPUT chain. In Fig. \ref{contourMaps}(a) we show the contour plots of the kinetic energy of the system versus space and time with lighter regions representing higher kinetic energy. The HS and the RF can be seen in Fig. \ref{contourMaps}(a). The simulations reveal that there are two kinds of high kinetic energy footprints in Fig. \ref{contourMaps}(a). We see fast moving (shown inside ellipses) and slow moving or nearly localized regions (shown inside rectangular boxes). Fig. \ref{contourMaps}(b) shows the contour map of just the HS in the chain. We should mention that localized excitations for significant times are rare in QEQ (see Ref. \cite{Ka17} for details). While the fast-moving excitations are identified as RF (shown inside ellipses), the semi-localized excitations (shown in rectangular boxes) are rejected based on our definition. {\it We also note that the RF we report here appear to be ubiquitous for all strongly nonlinear systems and appear to be distinct from the rogue waves alluded to in studies of Peregrine solitons of the {\it weakly} nonlinear Schr\"odinger equation} \cite{Pe83}. We further note here that while there is extensive evidence of existence of rogue waves, the oceanographic analyses of the origins of these waves in the open ocean is very much an evolving subject \cite{Ta80,Ta07,Fe16}.

\begin{figure}
\centering
\includegraphics[scale=0.120]{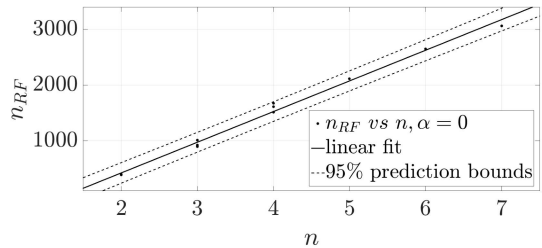}
\caption{Figure shows $n_{RF}$ as a function of $n$ and the linear fit for the data. Here $\alpha = 0$ and we have used $v_o = 0.58, 0.60$ and 0.63.}
\label{nRFVsN}
\end{figure} 

In earlier work on granular systems it has been shown that as $n\to\infty$, the width of a SW should shrink \cite{Su13}. We anticipate that with increasing $n$, the width of the SWs would be smaller and hence the number SWs that can be accommodated in a system ought to increase. Such an increase is expected in due course to increase the number of RF in the system. This can be seen clearly in Fig. \ref{nRFVsN} for the case $\alpha = 0$. 

For Fig. \ref{nRFVsN} we have used a linear fit to the data obtained from the dynamical simulations,
\begin{eqnarray}
n_{RF} & = & an + b,
\end{eqnarray}
where $n_{RF}$ is the total number of RF found in the system. The parameters of the fit are calculated to be $a = 550\pm 32$ and $b = -680\pm 130$. We observe that this growth behavior of RF with increasing $n$ for $1 < n \leq 1.3$ in the Hertz potential (i.e., greater than quadratic and less than quartic) is exponential in nature as reported in our earlier work \cite{Ha14}. The $1.3 < n < 2$ region is not readily accessible in the studies reported here for the $\beta$-FPUT chain. The nonlinear regimes with $1 < n <  2$ and the $n \ge 2$ regimes cannot be easily connected in this work. However, our earlier study in Ref. \cite{Ha14} and the current work together {\it suggest} that strongly nonlinear systems are generically prone to RF in QEQ. 

\begin{figure}
\centering
\includegraphics[scale=0.300]{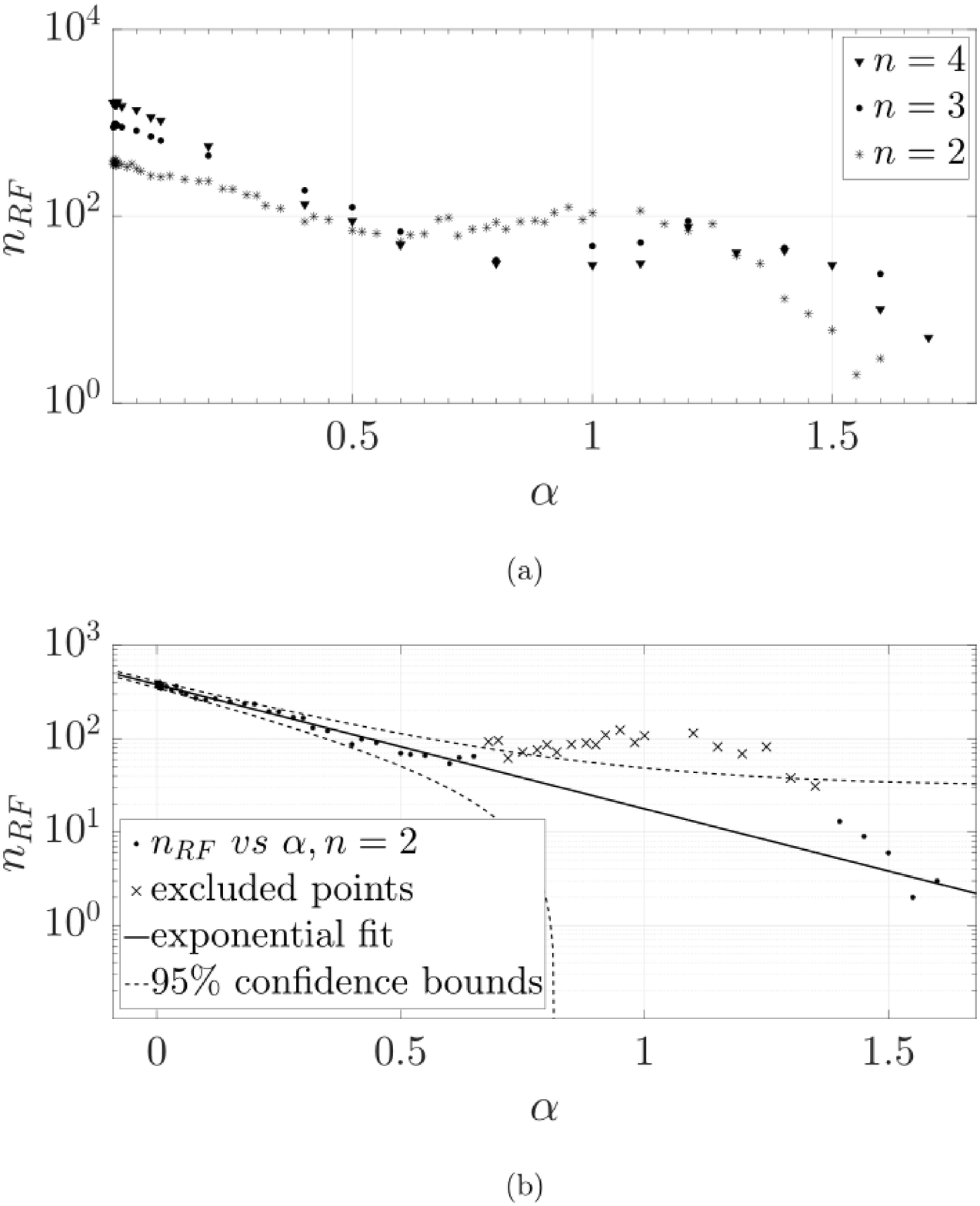}
\caption{Subfigure(a) shows $n_{RF}$ as a function of $\alpha$ for $n = 2, 3$ and $4$ respectively. Subfigure (b) shows the exponential fits to the data, while excluding the co-existence regime. Here, $v_o = 0.60$.}
\label{nrfVsAlphaFits}
\end{figure}

Harmonic forces can be introduced in the system by setting $\alpha > 0$. Typically, harmonic oscillations tend to progressively disperse SWs in a system \cite{Fl89,Se03,We18,Re02}. Therefore, we expect that increasing $\alpha$ would decrease $n_{RF}$. Our dynamical simulations strongly suggest that $n_{RF}$ decreases exponentially with increasing $\alpha$ as expected except for a region of $\alpha$ where $n_{RF}$ shows {\it an unexpected rise followed by a fall to the exponential decay with increasing} $\alpha$ as can be seen in Fig. \ref{nrfVsAlphaFits}. 

Initially this unexpected increase in $n_{RF}$ over a window of increasing $\alpha$ may seem like an error. However, after extensive analyses of the results such as exploring whether there are errors in energy conservation over extended time simulations and repeating the calculations with slightly changed parameters we found that the effect showed up in every study and hence is {\it real}. What is remarkable is that the behavior is observed for all values of $n$ that we are able to explore while maintaining the energy conservation accuracy over long time simulations. 

The simulations also suggest that the maximum number of RF are realized for progressively larger values of $\alpha$ as $n$ increases (see Fig. \ref{nrfVsAlphaFits}(a)). In earlier work, we have reported strongly nonlinear behavior when the linear and nonlinear parts of the potential become highly competitive leading to exceedingly long-lived system dynamics and absence of relaxation \cite{Ta12,Jo90,Av20,Wu20}. Further, the system behaves almost like an integrable system \cite{Ne01} with the SW-SW interactions being much weaker than what is seen in the $\beta$-FPUT system \cite{Wu20}. Hence, in retrospective, the observed behavior is perhaps not entirely unexpected. 

In summary, our studies suggest that the system behaves as a {\it nearly integrable system} in this special regime where the acoustic oscillations and the solitary waves become weakly interactive. This is so possibly because the length scales and the velocities associated wth these typical acoustic-like waves is comparable to that of the SWs \cite{Ta12,Jo90,Av20,Wu20}.   

Excluding the co-existence regime, we find that the exponential function provides a suitable fit as shown in Fig. \ref{nrfVsAlphaFits} (b) for $n = 2$. The fit uses the following function
\begin{eqnarray}
n_{RF} & \sim & e^{-\rho\alpha}.
\end{eqnarray}
We obtain $\rho = 3.07\pm 0.49, 3.90\pm 0.35$ and $5.1\pm 0.46$ for $n = 2,~3$ and 4, respectively. Interestingly, our fits suggest that $\rho \approx (n+1)$. Therefore, for progressively increasing $\alpha$, the RF get increasingly suppressed as $n$ increases.  The results in Ref. \cite{Ha14} were for various values of $v_o$ rather than for various values of $n$ as discussed here.

\subsection{Role of Initial Conditions}

\begin{figure}
\centering
\includegraphics[scale=0.300]{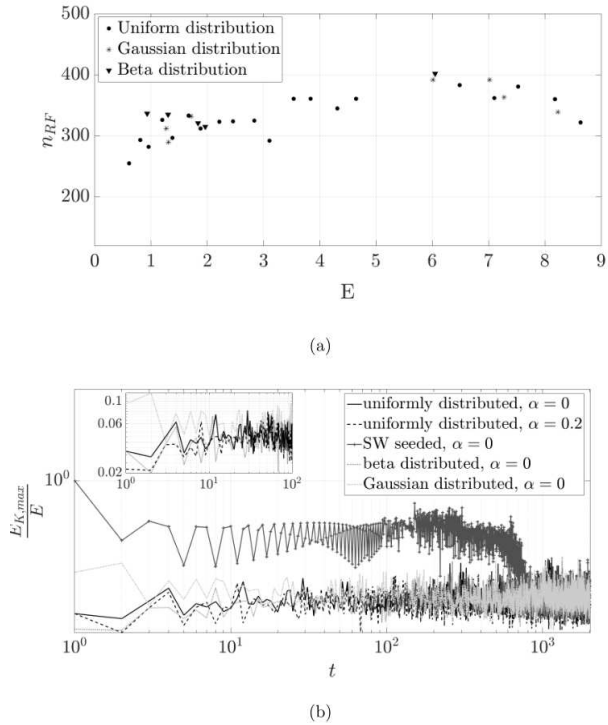}
\caption{Subfigure (a) shows $n_{RF}$ as a function of total energy of the system $E$, for three different distributions of the initial conditions - uniform, Gaussian and the beta distribution. Subfigure (b) shows the kinetic energy of the most energetic SW $E_{K,max}$, normalized by the total system energy $E$, for a range of initial conditions and system parameters. The inset shows early time transition of the system to QEQ for all the cases except the case of the single seeded SW. Both the plots show data for $\alpha = 0$ and $n = 2$.}
\label{nrfVsE}
\end{figure}

We now address the role of initial conditions in the formation of RF. In QEQ the system loses memory of its initial conditions \cite{Se06,Zw65,Be66,Be71}. The formation of RF, therefore, must not depend on the initial conditions. While we have used the uniform random distribution of particle velocities as an initial condition for the results reported discussed in this paper until now, we explored the cases where the initial velocities of the particles have been drawn from the beta and Gaussian distributions to test if our expectation was correct. Fig. \ref{nrfVsE}(a) shows $n_{RF}$ plotted against total system energy $E$ for various initial conditions with random velocities. In Fig. \ref{nrfVsE} (b), the dependence of the initial conditions on the onset of QEQ is explored. 

When the system is initiated by a single SW we see that the system eventually reaches the QEQ phase characterized by the small values of $E_{Kmax}/E$ (see the caption of Fig. \ref{nrfVsE}(b)) where RF are eventually observed. We thus show that the QEQ phase is reached regardless of the initial conditions used and this happens to be the case for the $\alpha = 0$ and $\alpha > 0$ cases with $n=2$ as shown in Fig. \ref{nrfVsE} (b). As we shall see below, many more interactions are needed when a SW is seeded at $t=0$ as opposed to when some random distribution of velocities is used to seed the dynamics. This explains why the case where the SW is seeded takes nearly a decade longer to reach the QEQ phase as seen in Fig. \ref{nrfVsE} (b). RF are seen in all the cases we have probed. 

The beta distribution used to explore a case of an initial random distribution of velocities is given by 
\begin{eqnarray}
P(x ; \alpha_D, \beta_D) & = & \frac{1}{B(\alpha_D, \beta_D)}x^{\alpha_D - 1}(1 - x)^{\beta_D - 1},
\end{eqnarray}
where, $B(\alpha_D, \beta_D)$ is a normalization constant. The parameters $\alpha_D$ and $\beta_D$ change the shape of the probability distribution. The Gaussian distribution which is also used to explore a separate case of random distribution of initial velocities is given by
\begin{eqnarray}
P(x ; \mu, \sigma) & = & \frac{1}{\sigma\sqrt{2\pi}}\exp\Big (-\frac{1}{2}\Big ( \frac{x - \mu}{\sigma} \Big )^2 \Big ),
\end{eqnarray}
where, $\mu$ and $\sigma$ are the mean and the standard deviation of the distribution, respectively. Although we have studied several cases of beta and Gaussian distributions for our studies, we show results from a selected study where we set $(\alpha_D, \beta_D) = (0.5, 0.5)$ and $(\mu, \sigma) = (0, 0.01)$ for the simulations corresponding to the beta and Gaussian distributions. For the two simulations with initial velocities sampled from the uniform distribution, we have set $v_0 = 0.6$. For the simulation with a single SW, we have set the initial velocity of particle 50 in the chain to 0.6 while the rest of the particles in the chain are unperturbed. As expected, we observe that the occurrence of RF is independent of initial conditions. Further, the occurrence of RF does not depend on the total energy of the system in a significant manner as seen in Fig. 5 (a).

We note that the number of RF we see in the studies shown here are $\sim 10^2$ in simulations across $10^{11}$ time steps. A question that will be addressed in future work has to do with the minimal conditions needed to realize RF in these strongly nonlinear systems?  RF are rare events and to understand {\it how rare and why} are daunting challenges for future analyses. Our preliminary studies along these lines suggest that the birth of RF presumably requires 5-6 energetic SWs to come together within a small time window and it is necessary to understand what conditions control such conditions.

\subsection{The RF in the Late QEQ Phase}

It is important to examine whether the RF recorded are late in the QEQ phase (as shown in Figs. \ref{contourMaps}) or past the same into an energy-equipartitioned state. Because if indeed RF are seen in the energy equipartitioned state, it could be possible to see them when systems are in equilibrium as well. 

In earlier work we have studied how the $\beta$-FPUT system with $n=2, \alpha=0$ at late enough times relaxes {beyond QEQ} to an energy equipartitioned state \cite{Ka19}. A key test used to determine the establishment of the equipartitioned state was to examine whether the simulations yielded a value of $C_v$ that can only be obtained when equilibrium prevails. Hence, it is reasonable to examine whether the system in our study relaxes past the QEQ state to an energy equipartitoned state by calcuating the specific heat $C_v$ across various time windows of relaxation. The key idea here is that if the equilibrium $C_v$ is attained then the system must have gone past the QEQ phase to the equilibrium phase. We followed the approach outlined recently by Przedborski {\it et al.} and presented in Eq. (16) in Ref. \cite{Pr17a} to carry out our calculations. 

For $n = 2, 3$ and 4, the theoretical values of $C_v$ when the systems reach the energy equipartitioned state were calculated to be 0.725, 0.64 and 0.608, respectively \cite{Pr17a}. The corresponding $C_v$ values from our simulations using Eq. (24) in Ref. \cite{Ka19} were found to be 0.741, 0.657 and 0.615. While these values are close to the values in the energy equipartitioned state, our experiences with these calculations (see \cite{Ka19}) suggest that they are not close enough to infer that the equipartitioned stage has been reached. Further, for the strongly nonlinear systems under various initial conditions being studied, the simulation times to reach the equipartitioned state turn out to be too long to simulate. Our results hence suggest that our study of RF is for systems that have {\it not} quite reached the equipartitioned state and that our calculations have been performed in QEQ. 

\section{Summary and Conclusions}\label{summary}

The story of RF in non-integrable 1D systems with strongly nonlinear interactions is an interesting as well as an evolving one. Given that RF emerge late into system dynamics, and that nonlinear dynamical simulations need to be done with high precision, it is challenging to resolve even relatively simple questions within a short time. Our studies in this work were done on RF as the systems evolved through the time steps outlined in Sec. \ref{sec:2}. Hence, in this second paper on RF, we lay out our progress and close with questions that we wish to address in future work.

We note here that depending on the nature of the interactions, these systems when subjected to a perturbation typically exchange energy between the particles using energy carriers that are characteristic of strongly nonlinear systems. Depending upon the details of the interactions, the carriers can be some or all of propagating SWs, localized nonlinear excitations, and objects that have SW and localized excitation-like features along with acoustic-like oscillations \cite{Ka17}.  Interactions between these objects are non-trivial. Eventually these systems evolve into the QEQ phase characterized by no memory of initial conditions, a Gaussian distribution of velocities and no equipartitioning of energy. Our contention is that RF seem to appear in the QEQ phase and possibly are characteristic of strongly nonlinear systems. 

In earlier work \cite{Ha14}, we showed that for granular chains held within fixed boundaries characterized by Hertz and Hertz-like potentials with $1 < n < 1.3$, $n_{RF} \sim \exp({\gamma}2n)$, where $\gamma = 6.41 \pm 0.81$. Here we report that for $\beta$-FPUT systems and similar systems with sextic, octic, etc potentials (see Eq. (2)), i.e., for $n \ge 2$, where $n$ is an integer, $n_{RF} \sim an + b$, where $a = 550 \pm 32$ and $b = -680 \pm 130$ are constants.  The region between $1.3 < n < 2.0$ where the cross-over happens from exponential to linear growth of $n_{RF}$ with respect to $n$ cannot be explored using the $\beta$-FPUT like systems and will be the subject of future work using the Hertz-like potential.

Our studies show that increasing $\alpha$ in the $\beta$-FPUT chain suppressed $n_{RF}$ as $n_{RF} \sim \exp(-\rho\alpha$), where $\rho \approx (n + 1)$. This result is similar to the suppression of $n_{RF}$ reported with increasing precompression in the Hertz and Hertz-like chains (see Fig. 3 in \cite{Ha14}). However, in the Hertz chain we found that $n_{RF}$ decayed with precompression in a way that seemed consistent with a double-exponential function. While the Hertz-like and FPUT potentials are different, and there is no reason to expect that the suppression of $n_{RF}$ in all models would be the same, it would be interesting and important to understand more about how $n_{RF}$ gets suppressed by the presence of harmonic interactions for various model systems. 

The co-existence regime is where the harmonic and nonlinear pieces of the potential become competitive as alluded to in Refs. \cite{Ta12} and in \cite{Wu20} and the system shows behavior akin to that of a strongly nonlinear system. In our $\beta$-FPUT-like systems we see a similar co-existence phase when the harmonic and nonlinear forces become competitive, i.e., for a range of values of $\alpha$ given $\beta = 1$ and the power of $n$ in Eq. (2). Our investigations in \cite{Ta12} suggest that the system dynamics in this co-existence phase is similar to that in an integrable system where the SW never gets destroyed. Typically, in this state, we find that the SW interacts very weakly with the background oscillations of the particles in the chain. We contend that the SW seen in this regime for the Hertz chain problem is best described by Nesterenko's solitary wave solution for the strongly nonlinear Hertz chain under weak loading \cite{Ne01}. It is conceivable that Nesterenko's solution could hold the key to a better understanding of the co-existence phase in the $\beta$-FPUT like system.

Future work may need to address the following outstanding questions - (1) exactly when do the RF appear in the relaxation process of a perturbed chain and how long do they last, and how does their number distribution change as the system evolves? (2) Do RF exist in QEQ phases only or do they appear infrequently in equilibrium as well? (3) Are RF special to 1D systems or can they happen in higher dimensions?

\section{Acknowledgments}
SS has been partially supported by a Fulbright-Nehru Academic and Professional Excellence R-Flex Fellowship while at IIESTS, India where a part of this work was completed. The authors declare that they have no conflict of interest associated with publishing this work.


\begin{thebibliography}{}
%
%


\bibitem{Re03} M. Remoissenet, Waves called solitons: concepts and experiments, 3rd ed., Springer, Berlin (2003).

\bibitem{Ja91} E.A. Jackson, Perspectives in Nonlinear Dynamics vols. 1 and 2, Cambridge University Press, Cambridge (1991).

\bibitem{Wh74} G.B. Whitham, Linear and Nonlinear Waves, Wiley, New York (1974).

\bibitem{Ch75} N.H. Christ and T.D. Lee, Quantum expansion of soliton solutions, Phys. Rev. D 12, 1606-1627 (1975).

\bibitem{To70} M. Toda, Waves in nonlinear lattice, Prog. Theor. Phys. Suppl. 45, 174-200 (1970).

\bibitem{To81} M. Toda, Theory of nonlinear lattices, Springer, Berlin (1981).

\bibitem{Fe82} W.E. Ferguson, Jr., H. Flaschka and D.W. McLaughlin, Nonlinear normal modes for the Toda chain, J. Comput. Phys. 45, 157-209 (1982).

\bibitem{Fo92} J. Ford, The Fermi-Pasta-Ulam problem: paradox turns discovery, Phys. Repts. 213, 271-310 (1992).

\bibitem{Sh14} Y. Shen, P.G. Kevrekidis, S. Sen and A. Hoffman, Characterizing traveling-wave collisions in granular chains starting from integrable limits: The case of the Korteweg--de Vries equation and the Toda lattice, Phys. Rev. E 90, 022905 (2014).

\bibitem{Ab76} K.O. Abdulloev, I.G. Bogolubsky and V.G. Makhanov, One more example of inelastic soliton interaction, Phys. Lett. A 56, 427-428 (1976).

\bibitem{Ei77} J.C. Eilbeck and G.R. McGuire, Numerical study of the regularized long wave equation II: Interaction of solitary waves, J. Comput. Phys. 23, 63-73 (1977).

\bibitem{Bo80} J.L. Bona, W.G. Pritchard and L.R. Scott, Solitary-wave interaction, Phys. Fluids 23, 438-441 (1980).

\bibitem{Na81} H. Nagashima and M. Kuwahara, Computer simulation of solitary waves of the nonlinear wave equation $u_t + uu_x - \gamma^2u_{5x} = 0$, J. Phys. Soc. Jpn. 50, 3792-3800 (1981).

\bibitem{Ko87} Y. Kodama, On solitary wave interactions, Phys. Lett. A 123, 276-282 (1987).

\bibitem{Se01} S. Sen and M. Manciu, Solitary wave dynamics in generalized Hertz chains: An improved solution of the equation of motion, Phys. Rev. E 64, 056605 (2001).

\bibitem{De20} G. Deng, G. Biondini and S. Sen, Interactions of solitary waves in integrable and nonintegrable lattices, Chaos: An Interdisciplinary Journal of Nonlinear Science. 30, 043101 (2020).

\bibitem{Bi78} A.R. Bishop and T. Schneider eds., Solitons and Condensed Matter Physics, Proceedings of a symposium on nonlinear (soliton) structure and dynamics in condensed matter, Oxford, June 27-29, 1978, Springer, Berlin (1978).

\bibitem{Ka17} R. Kashyap, A. Westley, A. Datta and S. Sen, Early time evolution of a localized nonlinear excitation in the $\beta$-FPUT chain, Int. J. of Mod. Phys. B 31, 1742014 (2017).

\bibitem{Fl89} N. Flytzanis, St. Pnevmatikos, M. Peyrard, Discrete lattice solitons: properties and stability, J. Phys. A: Math. Gen. 22, 783-801 (1989).

\bibitem{Se03} S. Sen, S. Chakravarti, D.P. Visco, Jr., M. Nakagawa, J. Agui, Jr. and D.T. Wu, Impulse propagation in granular systems, Proceedings of PASI on Modern Challenges in Statistical Mechanics, in V.M. Kenkre and K. Lindenberg eds., AIP Conf. Proc., vol. 658, pp. 357-379, Ridge, New York (2003)

\bibitem{KM09} S. Sen and T.R. Krishna Mohan, Dynamics of metastable breathers in nonlinear chains in acoustic vacuum, Phys. Rev. E 79, 036603 (2009).

\bibitem{Av11} E Avalos, D Sun, R.L. Doney, and S Sen,	Sustained strong fluctuations in a nonlinear chain at acoustic vacuum: Beyond equilibrium, Phys. Rev. E 84, 046610 (2011).

\bibitem{Ma00} M. Manciu, S. Sen and A.J. Hurd, Crossing of identical solitary waves in a chain of elastic beads, Phys. Rev. E 63, 016614 (2000).

\bibitem{Ma02} F.S. Manciu and S. Sen, Secondary solitary wave formation in generalized Hertz chains, Phys. Rev. E 66, pp. 016616 (2002).

\bibitem{Jo05} S. Job, F. Melo, A. Sokolow and S. Sen, How Hertzian solitary waves interact with boundaries in a 1D granular medium, Phys. Rev. Lett. 94, 178002 (2005).

\bibitem{Sa11} F. Santibanez, R. Munoz, A. Caussarieu, S. Job and F. Melo, Experimental evidence of solitary wave interaction in Hertzian chains, Phys. Rev. E 84, 026604 (2011).

\bibitem{Ka19} R. Kashyap and S. Sen, PULSEDYN - a dynamical simulation tool for studying strongly nonlinear chains, Comput. Phys. Commun. 239, 134- 149, (2019).


\bibitem{Pa73} R.K. Pathria, Statistical Mechanics, 2nd ed., Pergamon, Oxford (1973), pp 147-150.

\bibitem{Li00} J.A.S. Lima and A.R. Plastino, On the classical energy equipartition theorem. Braz. J. Phys., 30, 176-180 (2000). 

\bibitem{Lo86} S.W. Lovesey, Condensed Matter Physics: Dynamic Correlations, Addison-Wesley, Reading (1986).

\bibitem{Le82} M.H. Lee, Solutions of the generalized Langevin equation by a method of recurrence relations, Phys. Rev. B 26, 2547-2551 (1982).

\bibitem{Le83} M.H. Lee, Can the velocity autocorrelation function decay exponentially?, Phys. Rev. Lett. 51, 1227-1230 (1983).

\bibitem{Se91} S. Sen, Exact solution of the Heisenberg equation of motion for the surface spin in a semi-infinite s=1/2 XY chain at infinite temperatures, Phys. Rev. B 44, 7444-7450 (1991).

\bibitem {Se06} S. Sen, A tutorial on solving the Liouville equation: a formalism and an example, Physica A 360, 304-324 (2006).

\bibitem{Le07} M.H. Lee, Birkhoff's theorem, many-body response functions, and the ergodic condition, Phys. Rev. Lett. 98, 110403 (2007).

\bibitem{Ki46} J.G. Kirkwood, The statistical mechanical theory of transport processes: I. General theory, J Chem Phys 14, 180-201 (1946).

\bibitem{Zw65} R. Zwanzig, Time-correlation functions and transport coefficients in statistical mechanics, Ann. Revs. Chem. Phys. 16, 67-101 (1965).

\bibitem{Be66} B.J. Berne, J.P. Boon and S.A. Rice, On the calculation of autocorrelation functions of dynamical variables, J. Chem. Phys. 45, 1086-1096 (1966).

\bibitem{Be71} B.J. Berne and D. Forester, Topics in time dependent statistical mechanics, Ann. Revs. Chem. Phys. 22, 563-596 (1971).

\bibitem{Ku2} R. Kubo, M. Toda ad N. Hashitsume, Statistical Physics II: Nonequilibrium Statistical Mechanics, Springer, Berlin (1978).




\bibitem{Ha14} D. Han, M. Westley and S. Sen, Energy fluctuations in the granular chain: possibility of rogue fluctuations or waves, Phys. Rev. E 90, 032904 (2014).

\bibitem{He81} H. Hertz, \"Uber die Ber\"uhrung fester elastischer K\"orper, J f\"ur de reine u Angew. Mathematik 92, 156-171 (1881).

\bibitem{Se04} S. Sen, T.R. Krishna Mohan and J.M.M. Pfannes, The quasi-equilibrium phase in nonlinear 1D systems, Physica A 342, 336-343 (2004).

\bibitem{KM05} T.R. Krishna Mohan and S. Sen, A new equilibrium phase in discrete nonlinear chains, Pramana - J. Phys. 64, 423-431 (2005).


\bibitem{Se05} S. Sen, J.M.M. Pfannes and T.R. Krishna Mohan, The quasi-equilibrium state: a tale of certain soundless systems, J. Korean Phys. Soc. 46, 571-573 (2005).

\bibitem{Av09} E. Avalos and S. Sen, How solitary waves collide in discrete granular alignments, Phys. Rev. E 79, 046607 (2009).


\bibitem{Fu20} N.J. Fuller and S. Sen, Nonlinear normal modes in the $\beta$-Fermi-Pasta- Ulam-Tsingou chain, Physica A 553, 124283 (2020).



\bibitem{Pr17b} M. Przedborski, S. Sen and T. Harroun, The equilibrium phase in heterogeneous Hertzian chains, J. Stat. Mech: Theory and Experiment, 123204 (2017).


\bibitem{Pr17a} M. Przedborski, S. Sen and T. Harroun, Fluctuations in Hertz chains at equilibrium, Phys. Rev. E 95, 032903 (2017).

\bibitem{Be04} J. Berges, Sz. Borsanyi and C. Wetterich, Prethermalization, Phys. Rev. Lett. 93, 142002 (2004).



\bibitem{Gr12} M. Gring, M. Kuhnert, T. Langen, T. Kitagawa, B. Rauer, M. Schreitl, I. Mazets, D. Adu Smith, E. Demler J. Schmiedmayer, Relaxation and prethermalization in an isolated quantum system, Science 337, 1318-1322 (2012).

\bibitem{Fe55} E. Fermi, J. Pasta, S. Ulam and M. Tsingou, Studies of the nonlinear problems No. LA-1940, Los Alamos Scientific Lab., N. Mex. (1955).

\bibitem{Ku86} R. Kubo, Brownian motion and nonequilibrium statistical mechanics, Science 233, 330-334 (1986).

\bibitem{Ve67} L. Verlet, Computer ``experiments'' on classical fluids. I. Thermodynamical properties of Lennard-Jones molecules, Phys. Rev. 159, 98-103 (1967).

\bibitem{Al87} M.P. Allen and D.J. Tildesley, Computer simulation of liquids, Clarendon, Oxford (1987).

\bibitem{Ha92} J.M. Haile, Molecular Dynamics simulations: elementary methods, John Wiley, New York (1992).

\bibitem{Pe83} D.H. Peregrine, Water waves, nonlinear Schr\"odinger equations and their solutions, The ANZIAM Journal 25, 16-43 (1983).

\bibitem{Ta80} M.A. Tayfun, Narrow band nonlinear sea waves, J. Geophys. Res. 85, 1548-1552 (1980).

\bibitem{Ta07} M. Tayfun and F. Fedele, Wave height distributions and nonlinear effects, Ocean Engg. 34, 1631-1649 (2007).

\bibitem{Fe16} F. Fedele, J. Brennan, S. Ponce de Le\'on, J. Dudley and F. Dias, Real world ocean rogue waves explained without the modulational instability, Sci. Repts. 6, 27715 (2016).

\bibitem{Su13} D. Sun and S. Sen, Nonlinear grain-grain forces and the width of the solitary wave in granular chains: a numerical study, Gran. Matt. 15, 157-161 (2013).

\bibitem{We18} A. Westley and S. Sen, Solitary waves and localized excitations in the strongly nonlinear $\beta$-Fermi-Pasta-Ulam-Tsingou chain, Europhys. Lett., 123, 30005 (2018).


\bibitem{Re02} R. Reigada, A. Sarmiento and K. Lindenberg, Energy relaxation in Fermi-Pasta-Ulam arrays, Physica A 305, 467-485 (2002).

\bibitem{Ta12} Y. Takato and S.Sen, Long-lived solitary waves in granular chains, Europhys. Lett. 100, 24003 (2012).

\bibitem{Jo90} J.G.H. Joosten, E.T.F. Gelad\'e and P.N. Pusey, Dynamic light scattering by nonergodic media: Brownian particles trapped in polyacrylamide gels, Phys. Rev. A 42, 2161-2175 (1990).

\bibitem{Av20} E. Avalos, A. Datta, A.D. Rosato, D. Blackmore and S. Sen, Dynamics in a confined mass-spring chain with $1/r$ repulsive potential: Strongly nonlinear regime, Physica A 553, 124651 (2020).

\bibitem{Wu20} Q. Wu, X. Liu, T. Jiao, S. Sen and D. Huang, Head-on collision of solitary waves described by Toda lattice model in granular chain, Chin. Phys. Lett. 37, 074501 (2020).

\bibitem{Ne01} V. Nesterenko, Dynamics of heterogeneous materials, Springer, New York (2001), Chapter 1, Sec. 4.

%
%
%
%


\end{thebibliography}
\end{document}